\documentclass[pdflatex,sn-mathphys-num]{sn-jnl}

\usepackage{capt-of}
\usepackage{balance,bm}
\usepackage{color, colortbl, xcolor}

\usepackage{url}
\usepackage{hyperref}
\hypersetup{
    colorlinks=true,
}

\usepackage{subcaption}
\usepackage{booktabs} 
\usepackage{graphicx}
\usepackage{textcomp}
\usepackage{color,soul}
\usepackage{bm}
\usepackage{multirow}
\usepackage{wrapfig}
\usepackage{balance} 
\usepackage{fontawesome}
\usepackage[most]{tcolorbox}
\usepackage{graphicx}  
\usepackage{enumitem}
\usepackage{longtable}
\usepackage{colortbl}
\usepackage{arydshln}
\usepackage [english]{babel}
\usepackage [autostyle, english = american]{csquotes}
\definecolor{linkColor}{RGB}{6,125,233}
\definecolor{green}{rgb}{0.0, 0.65, 0.31}
\definecolor{bleudefrance}{rgb}{0.19, 0.55, 0.91}
\definecolor{ceruleanblue}{rgb}{0.16, 0.32, 0.75}
\definecolor{grey}{HTML}{969696}
\definecolor{violet}{HTML}{756bb1}
\definecolor{dgrey}{HTML}{01665e}
\definecolor{lgrey}{HTML}{5ab4ac}
\definecolor{dgreen}{HTML}{005a32}
\definecolor{purple}{HTML}{ae017e}

\definecolor{editCol}{HTML}{000000}
\definecolor{maskCol}{HTML}{c51b7d}
\definecolor{lrColor}{HTML}{8856a7}
\definecolor{trColor}{HTML}{d01c8b}
\definecolor{ctColor}{HTML}{4dac26}
\definecolor{brickred}{HTML}{f03b20}
\definecolor{improveCol}{HTML}{253494}
\definecolor{worsenCol}{HTML}{d7191c}
\definecolor{DarkBlue}{HTML}{00008B}
\definecolor{mscolor}{HTML}{01665e}
\definecolor{nmscolor}{HTML}{bf812d}
\definecolor{lgreen}{HTML}{ccece6}
\definecolor{dolive}{HTML}{308014}

\definecolor{trColor}{HTML}{d01c8b}
\definecolor{ctColor}{HTML}{4dac26}
\definecolor{brickred}{HTML}{f03b20}
\definecolor{improveCol}{HTML}{253494}
\definecolor{worsenCol}{HTML}{d7191c}
\definecolor{lgreen}{HTML}{e0f3db}
\definecolor{dpink}{HTML}{CD1076}
\definecolor{pink}{HTML}{FED2D2}
\definecolor{soothinggreen}{HTML}{4dac26}
\definecolor{darkred}{HTML}{8B0000}

\definecolor{dblue}{HTML}{104E8B}
\definecolor{violet}{HTML}{8A2BE2}
\definecolor{mscolor}{HTML}{01665e}
\definecolor{nmscolor}{HTML}{d8b365}
\definecolor{deepgrey}{HTML}{525252}
\definecolor{dslate}{HTML}{2F4F4F}
\definecolor{dolive}{HTML}{556B2F}
\definecolor{teal}{HTML}{388E8E}
\definecolor{mscolor}{HTML}{01665e}
\definecolor{nmscolor}{HTML}{d8b365}

\definecolor{aicolor}{HTML}{018571}
\definecolor{occolor}{HTML}{ff7799}

\definecolor{srcolor}{HTML}{e34a33}
\definecolor{smcolor}{HTML}{253494}
\definecolor{srsmcolor}{HTML}{7fcdbb}
\definecolor{bothcolor}{HTML}{fe9929}
\definecolor{onecolor}{HTML}{018571}
\definecolor{marroon}{HTML}{881c1c}

\usepackage{mathtools}

\usepackage{amsmath}
\usepackage{array}
\usepackage{xcolor}
\usepackage{arydshln}
\usepackage{siunitx}
\colorlet{tablerowcolor4}{gray!50} 

\newcommand*{\textlabel}[2]{%
  \edef\@currentlabel{#1}
  \phantomsection
  #1\label{#2}
}
\usepackage{tcolorbox}

\colorlet{tableheadcolor}{gray!25} 
\colorlet{tablerowcolor}{gray!15} 
\colorlet{tablerowcolor2}{gray!45} 
\colorlet{tablerowcolor3}{gray!25} 

\newcommand{\rowcollight}{\rowcolor{tablerowcolor3}} %

\newif{\ifhidecomments}
  \hidecommentsfalse 
\ifhidecomments
    \newcommand{\keran}[1]{}
    \newcommand{\melissa}[1]{}
    \newcommand{\dongwhi}[1]{}
    \newcommand{\koustuv}[1]{}
    \newcommand{\ravi}[1]{}
\else
    \newcommand{\keran}[1]{\textbf{\small\sffamily{\textcolor{DarkBlue}{[#1 -- Keran]}}}}
    \newcommand{\melissa}[1]{\textbf{\small\sffamily{\textcolor{dolive}{[#1 -- Melissa]}}}}
    \newcommand{\ravi}[1]{\textbf{\small\sffamily{\textcolor{marroon}{[#1 -- Ravi]}}}}
    \newcommand{\dongwhi}[1]{\textbf{\small\sffamily{\textcolor{dpink}{[#1 -- Dong Whi]}}}}
    \newcommand{\koustuv}[1]{\textbf{\small\sffamily{\textcolor{violet}{[#1 -- Koustuv]}}}}
  \fi
\newcommand{\edit}[1]{{\textcolor{editCol}{#1}}}

\colorlet{tableheadcolor}{gray!25} 
\colorlet{tablerowcolor}{gray!5} 

\definecolor{neutralCol}{HTML}{dd1c77}
\definecolor{neutralGreen}{HTML}{31a354}
\definecolor{NewBlue}{HTML}{1879ba}
\definecolor{bleudefrance}{rgb}{0.19, 0.55, 0.91}  
\definecolor{AfTrColor}{HTML}{0868ac}  
\definecolor{BfTrColor}{HTML}{a8ddb5}  

\definecolor{AfCtColor}{HTML}{b10026}  
\definecolor{BfCtColor}{HTML}{fd8d3c}

\graphicspath{ {figures/} }

\newcommand{\para}[1]{\vspace{0.3em}\noindent\textbf{#1}~}

\newtcolorbox{takeaway}{
  enhanced,
  colback=white,
  colframe=blue!75!black,
  boxrule=0pt,
  frame hidden,
  borderline west={3pt}{0pt}{blue!75!black},
  left=10pt,
  right=10pt,
  top=8pt,
  bottom=8pt,
  sharp corners,
  before skip=10pt,
  after skip=10pt
}

\usepackage{graphicx}%
\usepackage{multirow}%
\usepackage{amsmath,amssymb,amsfonts}%
\usepackage{amsthm}%
\usepackage{mathrsfs}%
\usepackage[title]{appendix}%
\usepackage{xcolor}%
\usepackage{textcomp}%
\usepackage{manyfoot}%
\usepackage{booktabs}%
\usepackage{algorithm}%
\usepackage{algorithmicx}%
\usepackage{algpseudocode}%
\usepackage{listings}%

\theoremstyle{thmstyleone}%
\theoremstyle{thmstyletwo}%

\theoremstyle{thmstylethree}%

\begin{document}


\title[A Taxonomy of Mental Health Needs and Technology for Alzheimer's and Dementia Caregivers]{A Taxonomy of Mental Health and Technology Needs for Alzheimer's and Dementia Caregivers}


\author[1]{\fnm{Keran} \sur{Wang}}\email{keranw2@illinois.edu}

\author[1]{\fnm{Drishti} \sur{Goel}}\email{drishti4@illinois.edu}

\author[1]{\fnm{Jiayue Melissa} \sur{Shi}}\email{mshi24@illinois.edu}

\author[2]{\fnm{Violeta J.} \sur{Rodriguez}}\email{vjrodrig@illinois.edu}

\author[3]{\fnm{Daniel S.} \sur{Brown}}\email{daniel.s.brown@osfhealthcare.org}

\author[4]{\fnm{Dong Whi} \sur{Yoo}}\email{dy22@iu.edu}

\author[5]{\fnm{Ravi} \sur{Karkar}}\email{rkarkar@umass.edu}

\author[1]{\fnm{Koustuv} \sur{Saha}}\email{ksaha2@illinois.edu}

\affil[1]{\orgdiv{Siebel School of Computing and Data Science}, \orgname{University of Illinois Urbana-Champaign}, \orgaddress{\city{Urbana}, \postcode{61801}, \state{IL}, \country{USA}}}

\affil[2]{\orgdiv{Department of Psychology}, \orgname{University of Illinois Urbana-Champaign}, \orgaddress{\city{Champaign}, \postcode{61820}, \state{IL}, \country{USA}}}

\affil[3]{\orgdiv{Illinois Neurological Institute}, \orgname{OSF HealthCare}, \orgaddress{\city{Peoria}, \postcode{61603}, \state{IL}, \country{USA}}}

\affil[4]{\orgdiv{Department of Human-Centered Computing}, \orgname{Indiana University Indianapolis}, \orgaddress{\city{Indianapolis}, \postcode{46202}, \state{IN}, \country{USA}}}

\affil[5]{\orgdiv{Manning College of Information and Computer Sciences}, \orgname{University of Massachusetts Amherst}, \orgaddress{\city{Amherst}, \postcode{01002}, \state{MA}, \country{USA}}}


\abstract{
Family members caring for individuals with Alzheimer's disease and related dementias (AD/ADRD) provide the foundation of long-term care worldwide. In 2023, more than 11 million U.S. family and friends contributed 18 billion hours of unpaid care, often at the cost of their own physical and mental health. These informal caregivers---also referred as the ``invisible second patients''---experience elevated rates of mental health problems. 
Yet research commonly reduces their complex psychosocial experiences to a single construct of caregiver burden, obscuring which specific needs are unmet or effectively supported. At the same time, digital and AI-enabled technologies are rapidly expanding, from smartphone apps and videoconferencing to sensor platforms and AI chatbots. 
However, the absence of shared frameworks across medicine, psychology, and technology research limits cumulative progress.
This study introduces a Caregiver Mental Health and Technology Taxonomy that systematically links AD/ADRD caregiver needs with corresponding classes of technology-based interventions. Drawing from an interdisciplinary literature review and two qualitative studies with caregivers, the taxonomy identifies mismatches between caregiver priorities and existing technological support, highlights under-served domains such as relational strain and compassion fatigue, and proposes design directions for adaptive, responsive systems. The framework offers a shared vocabulary to guide clinicians, researchers, and technology designers in developing more person-centered and clinically grounded innovation in dementia care.}




\maketitle
\section{Introduction}\label{section:intro}



Alzheimer's disease and related dementias (AD/ADRD) are growing faster than formal long‑term‑care systems can respond. In 2023, 11 Million U.S. family and friends provided 18 Billion hours of unpaid care---worth \$346.6 Billion but exacting steep mental, physical, and financial tolls~\cite{better2023alzheimer}. 
By the year 2030, the annual global cost of dementia is projected to rise to \$2.8 Trillion~\cite{world2021global}. Across income settings, these informal caregivers are routinely called the ``invisible second patients'' of dementia care~\cite{shim2021caregiving,sutcliffe2017caring,prince2012strain}. Meta‑analyses show that one‑third of AD/ADRD caregivers meet criteria for major depression, and up to 44\% for an anxiety disorder---rates far above age‑matched non‑caregivers~\cite{sallim2015prevalence,cooper2007systematic}. However, most reviews collapse diverse emotions into a single construct of ``burden,'' obscuring which psychosocial needs are (or are not) alleviated. 
For example, a recent meta-analysis of tele-health interventions for AD/ADRD caregivers reported only modest reductions in global burden and depressive symptoms (SMD$\approx -0.15$) and did not identify which specific psychosocial needs improved~\cite{zhu2024effects}, underscoring how current evidence still treats caregiver distress as a single, undifferentiated construct. 
The lack of a rigorously derived taxonomy of expectations, needs, and intervention strategies to support the mental health of informal AD/ADRD caregivers leaves clinicians and providers without a framework for individualized care, informaticians without decision logic for integrating caregiver data into electronic health records, and digital health designers without an effective guidance to build adaptive and responsive support systems~\citep{cunningham2019understanding}.

In particular, the proliferation of digital tools---
videoconferencing, sensor platforms, smartphone apps, and (most recently) generative artificial intelligence (AI) and large language model (LLM)-driven chatbots---have shown promise in newer avenues for caregiver support~\cite{griffiths2018tele,gaugler2021remote,than2023smartphone,shi2025mapping,saha2025ai,goel2026rubrix,goel2026inform}. 
Although more than 150 AD/ADRD‑related apps are present for use~\cite{ali2024mhealth,chien2024technology}, yet inconsistent labels across nursing, behavioral medicine, human-computer interaction (HCI), and AI stymie cumulative science~\cite{gaugler2017consistency,gitlin2020dementia}. 
Audits of app quality reveal significant misalignments: the vast majority of apps (83\%) provide only generic educational content; only about 20\% explicitly state that their content is evidence-based or has been developed with, or reviewed by, clinicians or researchers (e.g., by citing clinical guidelines or naming clinical/academic partners), leaving most apps with no clear indication of expert vetting; only 6\% offer any form of personalization, and first-generation chatbots tend to exhibit superficial empathy without offering escalation pathways to human professionals or crisis services for more serious concerns~\citep{zou2024mhealth,shi2025mapping}.


To address the above gap in the literature, we aim to build a taxonomy that explicitly maps distinct caregiver mental health concerns to technological interventions. 
Our study is guided by the following research questions:

\begin{enumerate}[align=left]
    \item[\textbf{RQ1:}] What caregiver mental health needs have been empirically identified, and how are these needs associated with \edit{informal} caregivers' psychological and functional outcomes? 
    
    \item[\textbf{RQ2:}] What classes of technologies currently address \edit{informal} caregivers' mental health needs, and where do gaps or misalignments remain between these technologies and caregivers' lived experiences?
    
\end{enumerate}

To answer these questions, we triangulated insights from a structured interdisciplinary review and two qualitative studies with AD/ADRD caregivers conducted by our team. 
The resulting taxonomy offers a shared vocabulary, highlights well‑served and under‑served domains, and provides an actionable roadmap for clinicians, researchers, technology builders, and policy makers to design, evaluate, and implement interventions, that more effectively support caregiver mental health.

This work makes three contributions.
(1) A taxonomy of caregiver mental health needs that disaggregates “burden” into tractable domains (e.g., anticipatory grief, anxiety, guilt, compassion fatigue, social isolation, relationship management, self-efficacy, self-care).
(2) A technology alignment matrix that maps these needs to monitoring tools, informational resources, social and emotional support technologies, task-management tools, and hybrid AI platforms, distinguishing strong versus emerging evidence.
(3) A triangulation synthesis that surfaces convergent, contradictory, and complementary findings across literature and interviews, and yields design and implementation guidance for stage-sensitive, safety-aware interventions.

\section{Study and Methods}

Toward our research questions, we triangulated findings from two sources---1) a structured literature review across medicine, psychology, nursing, and human-computer interaction (HCI), and 2) two recent qualitative studies conducted by our team with AD/ADRD caregivers. 
The combination of these sources allowed us to synthesize a taxonomy that maps empirically grounded caregiver mental health needs to existing and emerging technology classes, identifying both alignments and gaps in current interventions.


\subsection{Literature Review}

\begin{figure}[t]
    \centering
    \includegraphics[width=0.9\textwidth]{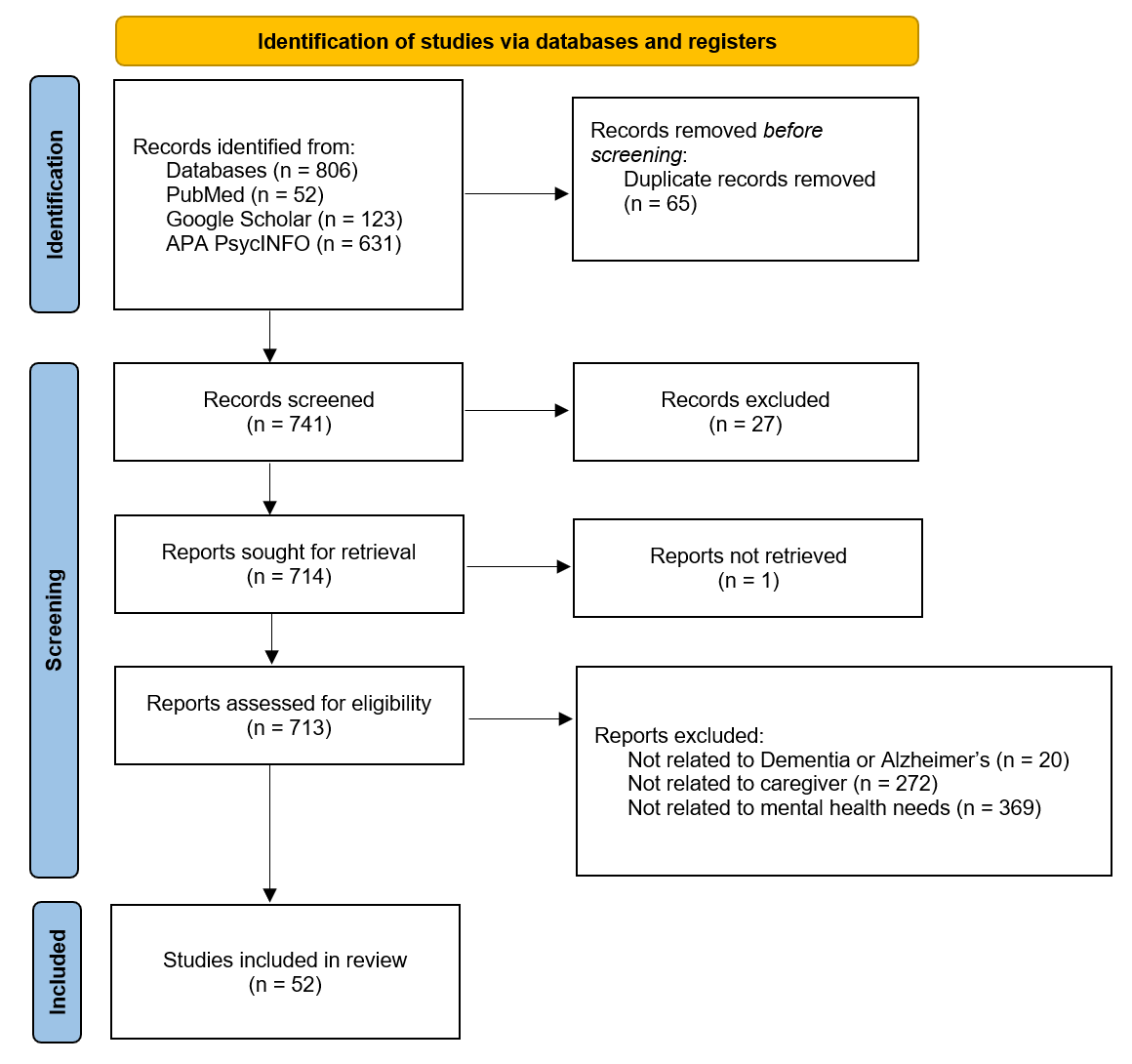} 
    \caption{PRISMA 2020 Flow Diagram of Study Selection}
    \label{fig:prisma}
\end{figure}

To conduct our literature review, we followed the Preferred Reporting Items for Systematic Reviews and Meta-Analyses (PRISMA) guidelines~\cite{page2021prisma} (ref: PRISMA flow diagram in~\autoref{fig:prisma}). 
\edit{Our study only considered English-language publications, as screening, data extraction, and synthesis were conducted in English.}

\para{Search.} PubMed, APA PsycInfo\textsuperscript{\textregistered}, and Google Scholar were queried (1 Jan 1975 – 10 Jun 2025) with the search phrases \textit{(``Alzheimer's'' OR ``dementia'') AND (``caregiver'') AND (``mental/ emotional/ psychological support needs'')}.

\begin{figure}[t]
    \centering
    \includegraphics[width=0.8\textwidth]{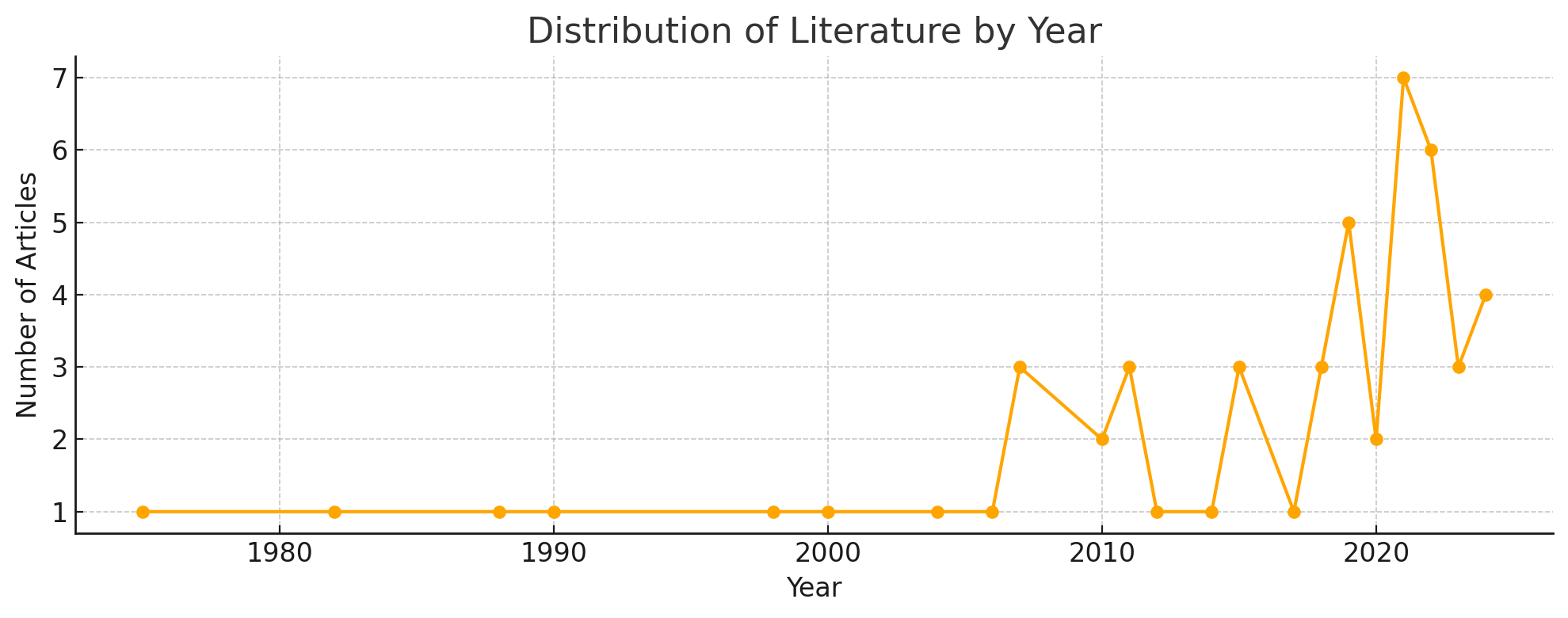}
    \caption{Publication year distribution of the included literature.}
    \label{fig:year-trend}
\end{figure}

\para{Selection process and study characteristics.} 
Screening proceeded in two stages. Titles and abstracts were first reviewed for obvious exclusions. Remaining articles underwent full-text appraisal against two \textit{a priori} criteria---1) \textbf{Population}: informal (family/friend) or formal (paid/professional) caregivers of individuals diagnosed with AD/ADRD, and 2) \textbf{Concept}: explicit discussion of caregiver mental health needs or concerns \textit{or} measurement of psychological outcomes arising from caregiving (e.g., burden, stress, anxiety, depression, grief).
Studies focusing solely on the person with dementia or mentioning mental health only incidentally were excluded, 
\edit{, i.e., when mental health was referenced only as background context and was not examined as a research question, measured outcome, intervention target, qualitative theme, or substantive finding.}
The application of these criteria identified 52 eligible studies for inclusion (see~\autoref{fig:year-trend} for a distribution of studies by year). These studies spanned both formal and informal caregiving contexts; however, consistent with the aims of this work, our analysis emphasizes findings related to informal caregivers. 

\begin{figure}[t]
    \centering
    \includegraphics[width=0.8\textwidth]{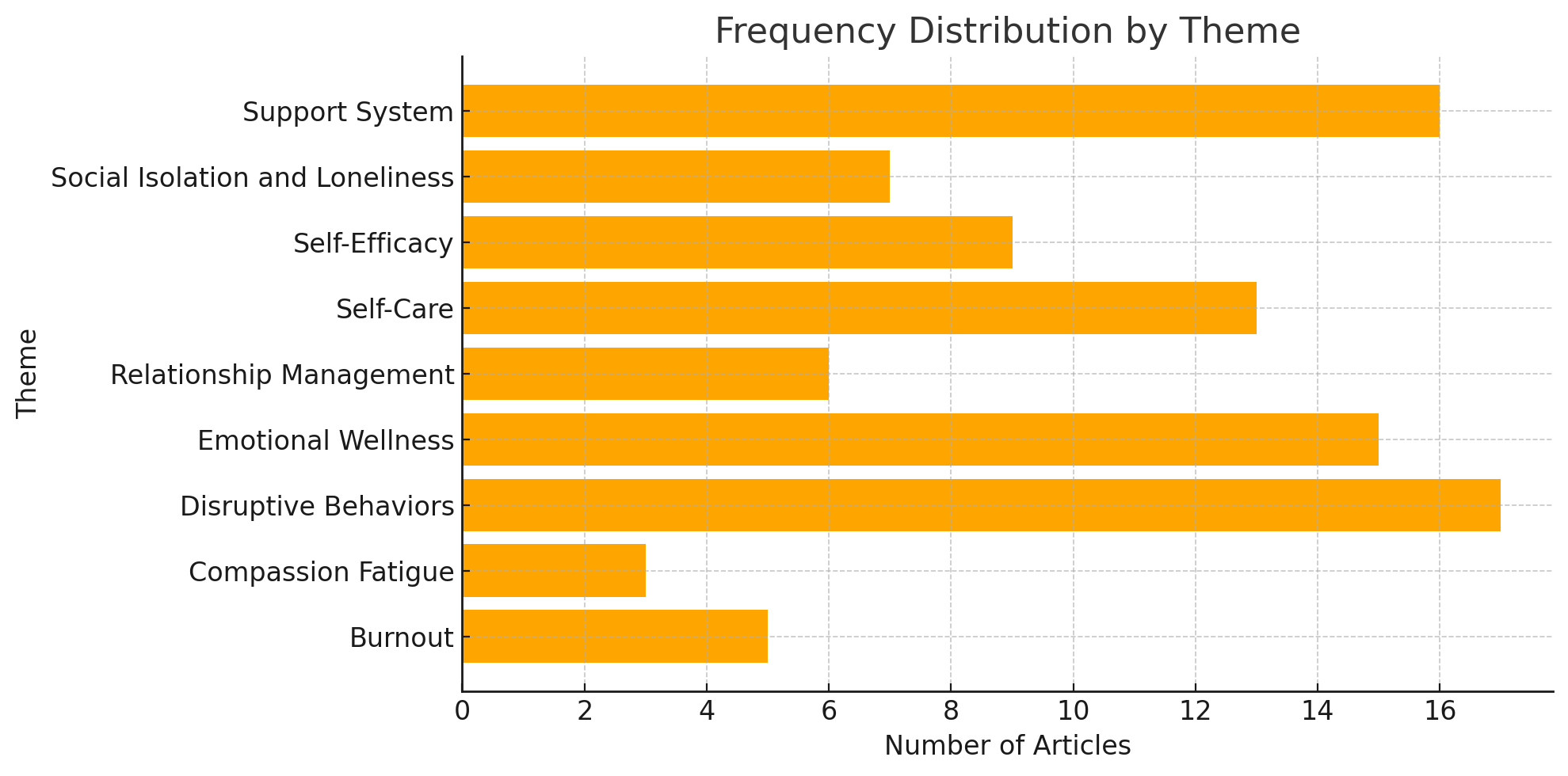}
    \caption{Distribution of themes in the literature on dementia caregiving.}
    \label{fig:theme-distribution}
\end{figure}

For each study, we logged mental health themes, key findings, study design and participant samples. We then applied narrative synthesis to cluster themes and identify evidence gaps.
The corpus of 52 eligible studies (1997–2025) comprised of 20 quantitative, 10 qualitative, 6 mixed‑methods, and 16 review papers; samples ranged from 8 to 26,701 participants (median=96).~\autoref{fig:theme-distribution} provides a distribution of themes of these studies.

\para{Best‑Fit Framework Synthesis.} 
To organize heterogeneous findings into a structured framework, we employed Best-Fit Framework Synthesis (BFFS)~\citep{carroll2013best}.
 As the a-priori scaffold, we adopted the widely used Alzheimer's Caregiver Stress-Process Model~\cite{pearlin1990caregiving}, which delineates the pathway from caregiving context to psychological outcomes through a series of stress domains. Specifically, the model includes---1) \textbf{Background \& Context}: sociodemographics, network size, programme availability, 2) \textbf{Primary Stressors}: cognitive decline, ADL/IADL dependence, disruptive behaviors, 3) \textbf{Secondary Role Strains}: family or work conflict, 4) \textbf{Financial Pressure}: social-life restriction, 5) \textbf{Secondary Intrapsychic Strains}: self-esteem, mastery, role-captivity, loss-of-self, and 6) \textbf{Outcomes}: with Coping and Social Support as cross-cutting mediators.


\subsection{Semi-structured Interview Studies}

\edit{We incorporated two qualitative studies conducted by our team to complement the literature review with first-hand caregiver accounts. While the literature review identified established constructs and intervention categories, the qualitative studies provided contextual detail on how these needs unfold in everyday caregiving and how caregivers interpret, adapt, or reject technology-based support. This combination allowed the taxonomy to reflect both the cumulative research evidence and the lived realities of informal AD/ADRD caregivers.}
To triangulate and refine this taxonomy, we incorporate findings from two recent qualitative studies, where our team conducted semi-structured interviews with 41 informal and family caregivers of individuals with AD/ADRD~\cite{shi2025balancing,shi2025mapping}. 
For both studies, interviews were audio/video recorded, then transcribed, anonymized, and analyzed using inductive coding and reflexive thematic analysis. 
The research team began with open coding of the transcripts, followed by affinity diagramming to collaboratively identify and refine key themes.
\edit{All interviews were conducted in English with participants who indicated comfort participating in English; no translation was involved, and interviewers used follow-up prompts and clarification questions to support accurate expression.}
These studies are briefly described below:

\subsubsection{Study 1: Mental health
needs of AD/ADRD family caregivers}  
Semi-structured interviews and thematic analyses explored everyday stressors, coping practices, and use of mental health technologies of AD/ADRD family caregivers ($N$=25)~\cite{shi2025balancing}. 
This study also captured the evolution of mental health concerns through the caregiving journey across the phases of early, middle, and late caregiving. 
Participants (P1.1--P1.25) engaged with eight literature-derived caregiving stressors, rating the severity of each on a 5-point Likert scale while thinking aloud to explain their reasoning and provide contextual insight.

\subsubsection{Study 2: Mapping Caregiver Needs to AI Chatbot Design} 
In this study, caregivers ($N$=16) engaged with a technology probe called Carey---an AI chatbot powered by GPT-4o~\cite{shi2025mapping}. 
The participants (P2.1--P2.16) shared experiences and perceptions about their interactions with Carey. 
Through semi-structured interviews and reflexive thematic analysis, the study surfaced caregivers' needs, expectations, and concerns regarding AI-based mental health support. These insights were then mapped to the current strengths and limitations of current AI capabilities, leading to design recommendations for building caregiver-centered AI tools for mental health and wellbeing. 

\section{Results}


The following sections present the triangulated results integrating insights from the interdisciplinary literature review and two qualitative studies with caregivers of individuals living with AD/ADRD. The findings are organized into two complementary components. First, we introduce a taxonomy of caregiver mental health needs, followed by an examination of the corresponding technology classes to evaluate how effectively current technological interventions address these needs. Within each thematic subsection, we synthesize evidence from both sources to identify where findings converge (literature and lived experiences align) and complement (interviews reveal additional nuances or contextual mechanisms). 
Collectively, our triangulation of findings
offer a coherent mapping of caregiver's psychosocial challenges to existing technological supports, clarifying which domains are well addressed, which remain partially served, and which are conceptually or practically underexplored.

\subsection{A Taxonomy of Caregiver Mental Health Needs}


The taxonomy presented in this section systematizes the primary needs of informal AD/ADRD caregivers. 
It expands upon eight empirically grounded domains---1) enduring disruptive care-recipient behavior, 2) navigating support systems, 3) building self-efficacy, 4) managing relationships, 5) preventing and mitigating compassion fatigue, 6) practicing self-care, 7) preventing burnout, and 8) regaining emotional balance.

\subsubsection{Enduring Disruptive Behaviors of the Care Recipients}
Living with AD/ADRD often exhibit a range of challenging behaviors beyond cognitive decline, including psychotic symptoms (delusions, hallucinations, confabulation), verbal and physical aggression, sleep-circadian disturbances, care resistance, and disorganized actions such as wandering or hoarding~\cite{swearer1988troublesome,cloak2019behavioral,baumgarten1990validity}. 
These behaviors not only complicate daily living assistance but also strain the caregiver-care-recipient relationship and evoke a sense of lost control~\cite{fauth2014behavioral,cheng2017dementia}. 
In fact, interpersonal tensions are often driven by delusions related to persecution, infedility, and misidentification~\cite{cheng2017dementia}.
These are clinically known as Behavioral and Psychological Symptoms of Dementia (BPSD), which demand constant 
supervision to prevent injury or getting lost, adding a significant burden on informal caregivers~\cite{miyamoto2010formal}. 
Additionally, nocturnal disturbances likewise disrupt caregiver sleep and wellbeing~\cite{tan2024psychological}.



\para{Convergent Findings.} Both literature and interviews identified verbal and physical aggression as major sources of distress. Prior studies link these behaviors to strained caregiver--care recipient relationships and loss of control~\cite{fauth2014behavioral, cheng2017dementia}. Across our interviews, caregivers described recurring disruptive-behavior clusters. Verbal aggression and outbursts were frequent and demoralizing (P1.3, P1.4, P1.7, P1.9, P1.10, P1.11, P1.17, P1.23); P3 reported \textit{``mostly\ldots{} verbal outburst, shouting at me using abusive language,''} feeling their sacrifice was unseen. Physical aggression ranged from episodes in memory care to assaults at home (P1.9, P1.10, P1.11, P1.18, P1.22, P1.23); P1.10 noted \textit{``100\% physical aggression\ldots{} she hit my sister,''} and P1.22 warned such outbursts could injure both parties. Repetitive occurrences included non-stop pacing and repeated demands to verify medication administration (P1.4, P1.10). Wandering, pacing, and restlessness---recognized in BPSD research~\cite{miyamoto2010formal}---were similarly reported (P4, P9, P10), reinforcing the constant supervision burden. Paranoia/delusional misinterpretations surfaced during acute phases (P18, P22), e.g., fears of being poisoned. Resistance or uncooperative expectations appeared as unrealistic task demands (P11). A milder cluster involved irritability or argumentativeness without aggression (P5, P13, P18); P13 rated concern ``2,'' with no hitting or throwing, and P5 described the recipient as \textit{``kind, very caring.''} Some reported minimal or absent disruptive behavior overall (P5 score~1; P12 score~2), underscoring heterogeneity. 

\para{Complementary Findings.} Interview data expanded the literature by revealing the emotional and relational toll of disruptive behaviors. P4 described \textit{``bossiness''} and \textit{``stalking people in the home,''} causing burnout and staff turnover, while P11 highlighted unrealistic demands that created frustration. These accounts reframed BPSD not only as behavioral management challenges but as drivers of compassion fatigue and emotional exhaustion.


\begin{takeaway}
\textbf{\faLightbulbO\ Takeaway:} Disruptive behaviors are not just ``tasks to manage''; they fundamentally reshape the relational and emotional climate of caregiving. Verbal and physical aggression, paranoia, wandering, and incessant demands erode predictability and induce a constant state of vigilance. Caregivers often interpret these behaviors as loss of reciprocity, loss of control, and loss of the person they knew. 
\end{takeaway}

\subsubsection{Efficient and Robust Support Systems}
A robust support system combines community resources, professional services, and formal care---such as senior centers, mental health professionals, home health aides, and respite programs---to alleviate caregiver stress and improve care quality~\cite{mueller2022systematic,queluz2020understanding}. 
Despite its importance, counseling and emotional support remain underutilized---many caregivers report unmet needs for psychological services, feel dismissed by providers who view dementia care as ``futile,'' and lack guidance on complex issues like end‐of‐life planning~\cite{zucca2022carers,zhang2020tensions,mclennon2021end,millenaar2018exploring}. Caregivers consistently express a desire for more comprehensive, empathetic advice from healthcare professionals.


\para{Convergent Findings.} Literature shows support exists but is under-used and hard to navigate~\cite{mueller2022systematic, queluz2020understanding, zucca2022carers}. Interviews mirror this invisibility and lack of guidance: \textit{``resources are built but people don't know they're there''} (P4, 2/5) and lamented that \textit{``no one tells you what's normal or what to expect,''} forcing families to figure things out (P10). Cost/time barriers observed in previous work~\cite{millenaar2018exploring, zhang2020tensions} were evident in practice---therapy was \textit{``very expensive\ldots{} you can't go many times''} (P3, 2/5); and paid employment constrained engagement with professional or community help (P12). Participants echoed documented frustrations with dismissed or inaccessible emotional care~\cite{zhang2020tensions, millenaar2018exploring}, describing a \textit{``major concern''} about finding responsive mental-health professionals or even \textit{``a nice medic''} (P6, P14), alongside fears of burnout and anxiety (P17, 5/5). Consistent with end-of-life guidance gaps~\cite{mclennon2021end}, caregivers reported minimal direction from primary care---\textit{``gives little,''}---and delayed discovery of useful advocates \textit{``much later than we should have''} (P4, P10).

\para{Complementary Findings.} Interviews extend the literature by highlighting the increasing centrality of informal and online peer communities in compensating for systemic gaps. Participants described Reddit and similar forums as \textit{``more helpful than therapy,''} valued for being free, empathetic and always available at home (P3, P11). They insights foreground a DIY ethos---\textit{``you have to do your own research''} (P18)--- where caregivers self-educate, and self-organize in response to fragmented formal systems. The data also surface relational asymmetries: families often remained attentive to the patient's wellbeing but neglect the caregiver's mental health (P22), while some found limited practical value in community agencies (P23). One participant with strong university-affiliated and local resources was \textit{``not at all concerned''} about support availability (P9). This exception underscores the contextual nature of sufficiency---where institutional proximity, socioeconomic factors, or community density can substantially alter caregiving experiences.



\begin{takeaway}
\textbf{\faLightbulbO\ Takeaway:} Caregivers face a dual burden---managing unpredictable behavioral escalations (verbal/physical aggression, wandering, perseveration) while navigating fragmented, ``poorly signposted'' support systems. In many cases, resources exist but remain inaccessible due to cost, time constraints, or the lack of guidance (``no one tells you what's normal''). Current technologies either address behavior management OR information access but rarely integrate both, leaving caregivers to build their own DIY support infrastructures.
\end{takeaway}

\subsubsection{Building Self-Efficacy}
Self‐efficacy---the belief in one's ability to handle caregiving challenges---influences goal setting, persistence, and coping strategies among AD/ADRD caregivers~\cite{bandura1997self}. Higher self‐efficacy correlates with greater recognition of caregiving's positive aspects (value, purpose, achievement), more effective emotion regulation in distressing situations, and lower depression alongside better physical health~\cite{semiatin2012relationship,khan2021self,gallagher2011self}. Cultural and gender factors also play a role---Hispanic/Latino caregivers often report higher self‐efficacy, possibly due to cultural values, whereas spousal caregivers---especially wives—may experience lower self‐efficacy linked to greater emotional strain~\cite{depp2005caregiver}.

\para{Convergent Findings.} Prior research consistently links self-efficacy to goal commitment, perseverance, and adaptive coping~\cite{bandura1997self, semiatin2012relationship}. The interview data aligned strongly with these themes, as caregivers described self-efficacy as strong, yet situationally fragile. Several voiced firm confidence in their competence---(\textit{``I don't doubt my ability\ldots{} I do it perfectly,''} (P3), \textit{``educated and prepared,''} (P5) and described effective in-the-moment coping (P10). Self-regulation following setbacks was common---P11's reminder that \textit{``good enough is good enough''} illustrates the internal reframing that helps sustain emotional equilibrium. Belief in one's capacity to handle difficult situations was linked to greater confidence and reduced anxiety (P17), echoing findings that self-efficacy buffers stress and promotes emotional regulation~\cite{semiatin2012relationship, gallagher2011self}.

\para{Complementary Findings.} Participants added value-laden motives that the literature seldom centers: P12 framed care as textit{``repaying''} her brother and P3 linked diligence to being \textit{``blessed,''} enriching accounts of positive appraisal and purpose~\cite{khan2021self}. The interviews also revealed social determinants of efficacy: P4 felt \textit{``failed''} when the family ignored recommendations and P13 struggled without clear benchmarks, extending efficacy beyond an individual belief to relational validation. While evidence often portrays self-efficacy as broadly protective~\cite{semiatin2012relationship, gallagher2011self}, several narratives revealed episodic fragility: pressures to do \textit{``more than what you are doing''} (P6), responsibilities \textit{``beyond my ability''} (P7), doubt triggered by persistent questioning (P18), and crisis-driven loss of focus and hope (P23).


\begin{takeaway}
\textbf{\faLightbulbO\ Takeaway:} Caregivers demonstrate confidence in their ability to provide care, but this confidence is highly situational. It is vulnerable to crisis episodes, family conflict, persistent questioning. It also carries moral and relational meaning. For many, caregiving was tied to duty, love, or identity, which could strengthen resolve but also make perceived failure more painful.
\end{takeaway}

\subsubsection{Managing Relationships}  
Primary caregivers often shoulder the majority of AD/ADRD‐related tasks, creating significant stress and intra‐family tension when siblings or other relatives contribute unevenly~\cite{smith2022family,katsarou2023investigating,smriti2024emotion}. 
Denial of the disease's progression or limited involvement by less‐engaged family members impairs dialogue and provokes disputes over care decisions, heightening the emotional burden on the lead caregiver~\cite{smith2022family,basnyat2021tensions}. In contrast, consistent emotional encouragement and practical assistance---such as shared meal preparation or joint attendance at medical appointments---from spouses or close kin have been shown to bolster resilience and reduce burnout risk~\cite{smith2022family,cleary2022interpersonal}. Cultivating open communication channels, establishing clear task assignments, and engaging all relatives in understanding the trajectory of the disease are, therefore, essential to balance patient needs with caregiver wellbeing and to preserve family cohesion.


\para{Convergent Findings.} Prior work links unequal task division among relatives to heightened stress and conflict~\cite{smith2022family, katsarou2023investigating}. Interviews mirrored this pattern vividly: caregivers often carried disproportionate responsibilities---\textit{``I'm the only one taking care of my mother\ldots{} I'm here struggling''} (P3); \textit{``it is a lot\ldots{} others who can help choose not to''} (P5)---while frustration frequently \textit{``bubbl[ed] up''} in family dynamics (P10). Some reported relatives who \textit{“walked away from us”} (P14), and in one case, a scheduling conflict led to a month-long estrangement (P10). These accounts align with literature, noting that denial, avoidance, and disengagement among family members often erode communication and amplify strain~\cite{smith2022family, basnyat2021tensions}.

\para{Complementary Findings.} Participants added nuance by emphasizing both stabilizing and destabilizing roles of close relationships. Consistent with findings that supportive partners buffer stress~\cite{smith2022family, cleary2022interpersonal}, several caregivers described maintaining equilibrium through \textit{``constant communication''} with family (P6). A caregiver reported \textit{``no tension''} as daughters divided tasks and visited regularly (P9), and some rated the relationship burden low amid stable cohesion (P11, P12, P13). Others illustrated the opposite outcome: lack of attention and emotional neglect led to a breakup and deepened isolation (P7). Some explicitly linked relationship quality to caregiving effectiveness noting that \textit{``strong, trusting relationships\ldots{} definitely improve the quality of care''} (P17).
Interviews also broadened the scope beyond kinship ties, capturing how caregiving reshapes social networks---\textit{``I don't have time to go out with friends''} (P3)---and complicates the balance between partners, siblings, and extended family (P23). Together, these accounts extend existing frameworks by situating family strain within broader social and relational ecosystems.


\begin{takeaway}
\textbf{\faLightbulbO\ Takeaway:} Family dynamics---especially uneven task distribution, denial of disease severity, disengaged siblings, or strained partnerships---are among the most prominent concerns which the caregivers linked to emotional distress. On the other hand, relational stability can be deeply protective: clear communication, equal divisions of tasks and emotionally present partners help alleviate strain. Relationship management is therefore not peripheral---it is a core determinant of caregiver wellbeing. 
\end{takeaway}

\subsubsection{Preventing and Mitigating Compassion Fatigue} \label{qual:compassion-fatigue} 
Compassion and empathy, while essential to caregiving, can exact high mental, physical, and economic costs~\cite{smriti2024emotion}. Sustained exposure to individuals experiencing trauma, distress, or progressive illness can adversely affect caregivers' psychological health, physical wellbeing, and overall quality of life---as well as that of their families, care recipients, and employing organizations. Compassion fatigue is a form of secondary traumatic stress that arises from repeated exposure to the suffering of traumatized individuals, leading to emotional, mental, and physical exhaustion and a diminished capacity for empathy~\cite{figley2013compassion}. It differs from occupational burnout in its emotional origins~\cite{wakefield2021future,flaubert2021supporting}. In AD/ADRD care, strong affective bonds amplify caregivers' awareness of relentless decline, fostering feelings of helplessness and grief that may culminate in neglectful behaviors or premature institutionalization of the care recipient~\cite{day2011compassion,schulz2007patient}. Structured respite services---such as scheduled adult day programs or in‑home relief---paired with peer support groups and targeted coping‑skills training in grief processing and boundary setting can interrupt this cycle, enabling caregivers to sustain empathy without becoming overwhelmed.


\para{Convergent Findings.} Literature defines compassion fatigue as a gradual loss of empathy and energy resulting from sustained exposure to suffering~\cite{wakefield2021future, flaubert2021supporting}. Interview narratives reflected this pattern: caregivers described prolonged emotional and physical exhaustion, feeling---\textit{``a weighted blanket\ldots{} your motivation just disappears''} (P04), \textit{``working 24/7 with almost no rest''} (P06), and living \textit{``life stuck in limbo''} (P04). Feelings of helplessness amid dementia's unending decline, consistent with prior findings linking prolonged strain to empathic erosion~\cite{day2011compassion, schulz2007patient}. Emotional closeness often amplified distress, as P11 acknowledged feeling \textit{``guilty even though I knew compassion fatigue is normal,''} and P07 continued caregiving because \textit{``you can’t stop caring.''} These accounts illustrate the tension between enduring attachment and the cumulative toll of unrelieved caregiving demands.

\para{Complementary Findings.} Interviews revealed concrete behavioral and coping markers---unhealthy cravings or drinking (P04), irritability towards aggressive behaviors (P10), a perceived decline in empathy---feeling \textit{``less compassionate than I normally would''} (P22) and guilt over reduced empathy (P11, P22). Despite exhaustion, none discontinued caregiving; several admitted to momentary \textit{``snapping''} under pressure (P07, P10) yet employed active regulation strategies, including taking short breaks (P07), cognitive reframing (\textit{``it's the disease, not her,''} P10), and seeking online peer support (P11). These complement literature-endorsed interventions such as respite and peer programs~\cite{day2011compassion}, highlighting compassion fatigue as both an emotional and behaviorally detectable process amenable to targeted support. On the other hand, P03 reported that affection itself appeared to sustain resilience, and duty shielded against exhaustion---\textit{``I can't get tired of her.''}. This suggests that under certain conditions, emotional attachment can serve as a sustaining force rather than a depleting one.


\begin{takeaway}
\textbf{\faLightbulbO\ Takeaway:} Irritability, reduced empathy, cravings, and guilt appear as early indicators of compassion fatigue. Even though emotional closeness often amplifies distress, caregivers rarely stop caring; instead, they continue while feeling progressively more strained.
\end{takeaway}

\subsubsection{Practicing Self‑Care}  
Deliberate self‑care practices are critical for safeguarding caregivers' physical and mental health, yet many neglect their own medical appointments and medication regimens, leading to chronic pain, insomnia, or depressive symptoms~\cite{wang2015impact,wang2019impact}. More than half of informal caregivers report insufficient time for rest or leisure, exacerbating fatigue and psychological distress~\cite{zucca2022carers}. Prioritizing restorative routines---such as regular sleep schedules and brief daily breaks---alongside participation in social or spiritual activities and adherence to personal health check‑ups promotes sustained well‑being. Incorporating emotional self‑monitoring techniques helps caregivers recognize early stress indicators and maintain healthy boundaries~\cite{waligora2019self}.


\para{Convergent Findings.} Literature shows that over half of caregivers lack time for rest, leading to fatigue and distress~\cite{zucca2022carers}, a pattern echoed in our interviews. Caregivers reported \textit{``no time for myself\ldots{} I can't even go out''} (P3) or being entirely absorbed in caregiving---\textit{``my whole focus is on my mom\ldots{} no time for me''} (P7)---with some spending \textit{``over 16 hours''} each day providing care (P23). Even moments of rest were fleeting: \textit{``15 minutes\ldots{} not as much time as I needed''} (P11). Several participants described a constant state of vigilance, with one caregiver explaining, \textit{``I'm on the lookout for him all the time,''} (P12). These experiences reinforce evidence that sustained time pressure and surveillance demands deplete energy and compromise wellbeing~\cite{wang2015impact, wang2019impact}.

\para{Complementary Findings.} Participants extended the literature by explicitly linking self-care to caregiving effectiveness and quality. As P6 noted, \textit{``You have to be sound to take good care of them''}, emphasizing self-maintenance as a moral and functional duty, while P17 framed it as key to preventing burnout (P17). Some caregivers illustrated concrete coping strategies despite 24/7 demands: P4 practiced meditation, journaling, and exercise, and P13 maintained social routines such as lunches and hobbies, remarking, \textit{``you gotta suck it up, buttercup.''} These accounts enrich general recommendations for restorative routines and self-awareness~\cite{waligora2019self}, portraying self-care not merely as leisure, but as intentional resilience work sustained through habit, discipline and meaning-making.


\begin{takeaway}
\textbf{\faLightbulbO\ Takeaway:} Caregivers recognize that maintaining their own wellbeing is essential to providing sustainable care, yet self-care is difficult to carve-out by constant vigilance, long daily-care-hours, and round-the-clock responsibility. Many describe having \textit{``no time''} for themselves, experiencing interrupted rest, and living in a near-constant state of watchfulness. Nevertheless, caregivers strive to incorporate brief restorative routines---whether moments of quiet, social connection, or personal health maintenance.
\end{takeaway}

\subsubsection{Preventing Burnout} 

Burnout Syndrome is defined as a state of perceived overload and emotional exhaustion arising from prolonged interpersonal stressors in the workplace, characterized by three key dimensions: emotional exhaustion (lack of energy and emotional depletion), depersonalization (indifferent or cynical attitude towards patients), and reduced personal accomplishment (viewing caregiving efforts as negative or ineffective)~\cite{yildizhan2019burden,vicente2015burnout}. These factors can diminish the quality of care, leading to early patient institutionalization, social isolation, prolonged stress, and increased biological vulnerability~\cite{alves2019burnout}. Consequently, caregivers are at heightened risk for mental and physical health issues, such as hypertension, elevated stress hormones, suppressed immune function, major depression, and fatigue 
\cite{valente2013autopercepccao}. Across regions, common risk factors for caregiver burden included care-recipient behavioral problems, spousal caregiving, provision of ADL support, and extended caregiving hours exceeding 21 per week~\cite{chan2021caregiving}. Patient-side neuropsychiatric symptoms (BPSD) are not uniform drivers: irritability/agitation, hallucinations, aberrant motor behavior, and especially delusions show the strongest ties to caregiver exhaustion, whereas apathy---though prevalent---tends to be less burnout-provoking~\cite{hiyoshi2018behavioral, truzzi2012burnout}. Access to allied-health services—such as physiotherapy, occupational therapy, and speech therapy—was found to mitigate caregiver burnout, particularly by buffering the impact of care-recipient factors like advanced age (85+), co-residence with children, ADL dependency, mood disturbances, and intensive caregiving demands~\cite{chan2021caregiving}.

\para{Convergent Findings.} Literature defines burnout as emotional exhaustion, depersonalization, and reduced personal accomplishment~\cite{yildizhan2019burden, vicente2015burnout}, typically emerging from prolonged caregiving demands~\cite{chan2021caregiving}. Interview data strongly reflected these hallmarks. Caregivers described relentless routines and emotional depletion---\textit{``same routine, day in, day out\ldots{} I don't have anyone to help me''} (P3)---and rated their condition as severe---\textit{``burned out and overwhelmed''} (P5). Fatigue often intertwined with relationship strain and role overload---\textit{``the workload is way too much\ldots{} poor relationship\ldots{} leads to the breakup''} (P7);\textit{``managing my dad's care\ldots{} two sons\ldots{} back at work\ldots{} I'm just tired''} (P10). Worsening symptoms created a feedback loop of exhaustion and declining resilience, as one caregiver noted: \textit{``as the dementia worsened\ldots{} lack of restorative sleep\ldots{} a vicious cycle\ldots{} burnout worse\ldots{} ability to cope worse''} (P11). Participants tied these experiences to care quality---\textit{``I become really emotionally exhausted, less empathetic\ldots{} ignoring burnout can lead to mistakes''} (P17)---echoing evidence that chronice stress impairs both health and caregiving performance~\cite{valente2013autopercepccao, alves2019burnout}

\para{Complementary Findings.} Participants deepened existing frameworks by situating burnout within the social and relational complexity of family care. They linked depletion to strained partnerships and role conflict---\textit{``poor relationship\ldots{} leads to the breakup''} (P7); \textit{``managing my dad's care\ldots{} two sons\ldots{} back at work\ldots{} I'm just tired''} (P10)---and described burnout as cyclic and cumulative rather than a discrete state. Episodic peaks were frequent---\textit{``heavy work week\ldots{} very stressful and overwhelming''} (P12)---yet some reported moderation or adaptation over time---textit{``I’m not burned out yet''} (P13); \textit{``it's been a balance\ldots{} according to your mental ability''} (P14). Direct depersonalization was seldom voiced; instead, reduced empathy appeared during strain (P17). Self-directed recovery practices---\textit{``create time for yourself to rest''} (P6)---emerged as primary coping strategies, while few described engaging formal support, highlighting informal strategies over formal supports that the literature identifies as protective~\cite{chan2021caregiving}. Reduced empathy (P17) and tunnel vision (P11) replaced overt depersonalization, adding nuance to how burnout manifests in family caregiving.

\begin{takeaway}
\textbf{\faLightbulbO\ Takeaway:} Caregiver mental health operates across three interacting domains: external stressors (behaviors and systems) $\Rightarrow$ relational resources (efficacy and family support) $\Rightarrow$ internal depletion (fatigue to crisis). When disruptive behaviors escalate and systems fail, caregivers increasingly rely on family support. As inequities strain these relational resources, emotional depletion intensifies. Current technological interventions primarily address the external and internal domains but largely overlook the relational---the critical mediating layer where targeted support could disrupt the progression toward burnout.
\end{takeaway}

\subsubsection{Regaining Emotional Balance}
Maintaining emotional wellbeing in the context of AD/ADRD caregiving involves navigating a complex and shifting emotional landscape, despite the inherent challenges of caregiving, which include feelings of strain, depression, guilt, and embarrassment. This process entails understanding and acknowledging one's emotions, developing coping strategies, and seeking support when needed. 
Prior work noted a number of emotional challenges that are experienced by AD/ADRD caregivers. 
1) \textbf{Stress and Anxiety.} Caregivers commonly experience fear and worry about disease progression, emergencies, and their capacity to manage crises~\cite{katsarou2023investigating}. Reluctance to discuss end-of-life planning further exacerbates anxiety. The demands of caregiving and unpredictable behaviors often lead to frustration when positive feedback is scarce, trapping caregivers in repetitive tasks~\cite{katsarou2023investigating}. Constant vigilance induces hypervigilance, sustaining elevated stress as caregivers maintain alertness to prevent harm~\cite{christian2023understanding}.
2) \textbf{Grief and Loss.} AD/ADRD's gradual decline triggers \textbf{anticipatory grief}, wherein caregivers mourn losses of identity and independence long before death~\cite{cheung2018anticipatory,mclennon2021end}. Anticipatory grief encompasses sorrow over relational and cognitive losses and, if unaddressed, may evolve into complicated grief after death~\cite{stevens2024mixed}. Even after caregiving ends, relief that suffering has ceased often coexists with deep sadness and emptiness~\cite{mclennon2021end}.
3) \textbf{Guilt and Self‑Blame.} Guilt arises from perceived inadequate actions, personal limitations, negative emotions toward care recipients, shifts in the caregiver-recipient relationship, neglected responsibilities, and external expectations~\cite{gallego2022feel}. Some caregivers employ distress-avoidance to suppress guilt, yet many report self-reproach for not doing enough and guilt for feeling relief when suffering ends~\cite{clemmensen2021know,mclennon2021end}.

\para{Convergent Findings.} Prior work highlights fear/worry, hypervigilance, strain, depression, guilt, and embarrassment in AD/ADRD caregiving~\cite{katsarou2023investigating, christian2023understanding}. The interviews echoed these, and some caregivers described living under constant emotional pressure: one \textit{``felt like my mom was worsening\ldots{} started crying\ldots{} couldn't concentrate''} (P3) another worried persistently when away---\textit{``What if something happened\ldots{} when I was leaving''} (P9)---and a third carried the enduring \textit{``knowledge\ldots{} it's gonna get worse''} (P13). Vigilance often extended into public spaces, producing anxiety during \textit{``tantrums''} (P14), and guilt about leaving the care recipient alone (P23). Many portrayed chronic strain and depletion---\textit{``this becomes the big energy drain''} (P4), caregiving felt \textit{``really demanding\ldots{} carrying her''} (P7), like a \textit{``continual low-grade fever''} with periodic spikes (P11). Anxiety and low mood were pervasive (\textit{``anxiety… depression\ldots{} most times,''} P6; slight depression tied to disclosure, P12), while guilt recurred in varied forms—about insufficient visiting (P9), taking breaks (P23), or perceived past shortfalls (P4). Embarrassment arose around public incidents (P14), self-disclosure (P12), or family members who \textit{``question my efficacy''} (P22); isolation compounded this strain (P3, P7). Collectively, these accounts reinforce the emotional burden described in the literature, portraying caregiving as a state of ongoing vigilance, worry, and self-reproach intertwined with exhaustion.

\para{Complementary Findings.} Interviews detailed how emotions are managed: deliberate compartmentalizing (P5) and functional numbing to avoid \textit{``feel[ing] everything at full volume''} (P11); intentional joy-seeking (\textit{``seek\ldots{} that which brings you joy,''} P4); task delegation to reduce burden (P9); pharmacologic support aiding \textit{``depression and guilt''} (P18); and an explicit mood$\rightarrow$performance link (\textit{``when my emotion is negative\ldots{} efficiency\ldots{} decrease[s],''} P17). These extend coping/support nodes emphasized in prior work~\cite{katsarou2023investigating, christian2023understanding, gallego2022feel, clemmensen2021know, mclennon2021end}. Moreover, while previous studies often emphasize frustration from repetitive tasks~\cite{katsarou2023investigating}, participants in this sample tended to frame these experiences as strain, guilt, or low mood rather than irritation. As P13 reflected, it felt like hitting a \textit{``brick wall\ldots{} can’t go far anymore.''} Similarly, anticipatory loss was expressed through resigned awareness—\textit{``knowledge\ldots{} it’s gonna get worse''} (P13)---rather than explicit grief, suggesting a quieter, more diffuse form of emotional adaptation. This reframing moderates the prominence of anticipatory grief relative to prior reports~\cite{cheung2018anticipatory, mclennon2021end, stevens2024mixed}.

\begin{takeaway}
    \textbf{\faLightbulbO\ Takeaway:} AD/ADRD caregiving often entails a continual negotiation of stress, anticipatory loss, guilt, and embarrassment, intensified by constant vigilance and emotional strain. Caregivers work to regain balance through strategies such as compartmentalizing, seeking moments of joy, delegating tasks, and drawing on social or professional support. This underscores emotional regulation as an ongoing, adaptive practice central to sustaining long-term caregiving.
\end{takeaway}

\subsection{Technologies to Support the Mental Health Needs of Caregivers}\label{sec:tech}


Having identified the major mental health needs of the AD/ADRD caregivers, we next identified the range of technologies from the literature that are aimed at supporting the caregivers' mental health. We categorize these technologies into---1) monitoring tools, 2) informational resources, 3) social and emotional support technologies, 4) task management tools, and 5) AI-driven chatbots. 


\subsubsection{Monitoring Tools}
\label{sec:monitoring-tools}
Monitoring tools help track various physiological and behavioral signals over time, through technologies such as sensors, trackers, and cameras. 
For instance, wrist actigraphy tracks sleep-wake and correlates with depression, burden, and stress, providing objective feedback for self-care~\cite{palmese2024wearable}. 
Multisensor ambient platforms (e.g., People Power Caregiver) combining door, motion, and bed sensors reduced six-month anxiety for engaged users vs controls~\cite{levenson2024evaluating}. 
GPS trackers lowered elopement stress, raised perceived independence, and reduced burden, delaying 24/7 supervision~\cite{doyle2024implementing}. Night-time cues linking bed-pressure mats with smart lighting/audio reduced depression and anxiety over 12 weeks and helped carers \textit{``sleep more peacefully''}~\cite{ault2020smart}. Prior research noted how videoconference monitoring adds psychosocial support; webcam-delivered ACT improved anxiety, stress, quality of life, burden, and predeath grief~\cite{han2025videoconference}. 
\citeauthor{shaik2025advancing} found that mattress, gait, and other ambient sensors enhanced safety while lessening caregiver stress~\cite{shaik2025advancing}.
We summarize the gaps and misalignments below:
\begin{itemize}
\item Sensing rarely becomes interventions building self-efficacy or relationship management.
\item Social-isolation safeguards are sparse; integrate peer support or telepresence.
\end{itemize}

\para{Convergent Findings.} Prior work shows that monitoring systems can improve safety, reduce anxiety, and promote caregiver rest and emotional regulation~\cite{levenson2024evaluating, shaik2025advancing}. Interviews reflected this closely. Caregivers layered diverse ``monitoring'' tools---location trackers, in-home sensors, and lightweight analytics---to extend supervision and reclaim moments of rest. For elopement and wayfinding, several relied on GPS/location services such as AirTags and Life360 to \textit{``track a loved one during walks''} and verify safety remotely (P13, P9). After episodes of aggression, families escalated to room cameras and home alarm setups to maintain vigilance (P22, P18), with some emphasizing that cameras were installed \textit{``with consent''} and sometimes paired with video calling for quick check-ins (P6). Nighttime use was particularly salient: monitoring helped \textit{``manage caregiving while ensuring rest,''} allowing caregivers to sleep or step away while still feeling in control (P9).
Monitoring extended beyond safety to support routine adherence and wellbeing. Several integrated reminders, visual cues, and smart devices for medication management and daily structure---Wi-Fi pill dispensers that \textit{``worked well for a while''} (P10), Alexa prompts, analog clocks, simplified phones, or music players and dashboard displays (P5), and smartwatches for timing doses (P7). Physiological and emotional tracking appeared on both sides of care: caregivers used wearables to monitor their own blood pressure and sleep (P10), emotion tracking apps (P20), and self-designed dashboards, combining Excel, Power BI, and virtual assistants for scheduling and coordination (P6). Telepresence tools, especially the Echo Show, enabled communication across sensory or physical barriers: one caregiver used it for lip-readable virtual visits with a parent, while another relied on Echo for calls; in one instance, Echo-based check-ins triggered timely help when a parent became unwell (P10).
Still, effectiveness varied with disease progression---some tools provided early benefit but later waned (\textit{``worked well for a while,''} P10), and others showed \textit{``varying success''} across bundles of aids (P5). Families described trade-offs between surveillance and trust, emphasizing the importance of consent and escalation only after behavioral crises (P6, P9, P13, P18, P22). These accounts confirm findings that sensing technologies support safety and rest~\cite{ault2020smart, doyle2024implementing}, but also reveal the complex emotional and ethical terrain of continuous monitoring.


\para{Complementary Findings.} Literature's call to integrate telepresence and peer support and turn sensing into actionable workflows is grounded by Echo-mediated accessibility and rapid response (P10) and by caregiver analytics and virtual-assistant scheduling (P6), suggesting concrete paths to bolster support systems, relationship management, self-care, and emotional wellness. Safety-first sensing~\cite{shaik2025advancing} complements strategic camera use after aggression (P22, P18), while night-time cueing~\cite{ault2020smart} aligns with explicit rest and recovery aims (P09) to help prevent compassion fatigue and burnout. Caregiver's creative repurposing---such as combining home sensors, trackers, and communication aids---suggests that monitoring is most effective when treated as a flexible, person-specific ecosystem rather than a fixed intervention. Despite these alignments, interviews exposed the fragility of sustained engagement. While controlled trials report consistent mood or sleep benefits over fixed durations (e.g., 12-week interventions~\cite{ault2020smart}; anxiety reduction for engaged users~\cite{levenson2024evaluating}), caregivers described fluctuating results---tools \textit{``worked well for a while''} (P10) and bundles produced \textit{``varying success''} (P05). Some technologies faltered as dementia progressed, others introduced new burdens of oversight or privacy management. While the literature notes that sensing rarely builds self-efficacy or relationship management and lacks anti-isolation safeguards, participants describe sensing as deeply embedded in emotional regulation and coping. Ad-hoc analytics and scheduling (P6), telepresence reassurance (P10), and emotion-tracking tools (P20) served as personalized extensions of self-management---underscoring both the adaptability and the vulnerability of technology-supported caregiving over time. 


\begin{takeaway}
    \textbf{\faLightbulbO\ Takeaway:} Monitoring tools offer caregiving support by enhancing safety, reducing anxiety, and enabling caregivers to reclaim rest through layered systems of sensors, trackers, reminders, and telepresence. However, their effectiveness often changes with the progression of the disease, and caregivers must continually navigate trade-offs between vigilance, privacy, consent, and usability. 
\end{takeaway}

\subsubsection{Informational Resources (e.g., educational platforms, online support forums)}\par \label{sec:informational-resources}
\smallskip
Online and peer platforms provide scalable knowledge and support but vary in reliability. In \textit{Forum.hr}, 48.2\% of posts sought guidance on symptoms, progression, and treatment~\cite{erdelez2019online}; on Weibo, caregivers frequently queried medications and alternatives~\cite{hao2020studying}. Where professional moderation is absent, misinformation risk increases~\cite{coulson2024examining}; scoping reviews also document accuracy gaps in user-generated health content for long-term conditions~\cite{treadgold2025quality}.

Asynchronous peer spaces (e.g., Reddit r/Alzheimers) leverage anonymity and 24/7 access to deliver informational and emotional support, buffering loneliness and normalising distress~\cite{pickett2024social,kaliappan2025online}. 
Moderated online groups add trained facilitators and closed-group norms; effective programs cultivate \textit{``non-judgemntal''} climates and strengthen networks~\cite{daynes2023online}. Structured learning further lifts competence: the Understanding Dementia MOOC increases knowledge and improves behaviour/medical management~\cite{eccleston2019building}; Tele-STAR (STAR-C-Telemedicine) reduces behavioural symptoms and caregiver reactivity while conveying emotional support~\cite{lindauer2018took}; the UK \textit{``Talking Point''} forum improved relationship quality (SQCRC, $P=0.003$) after 12 weeks~\cite{mckechnie2014effectiveness}; and FamTechCare's video-feedback coaching increased confidence and lowered perceived severity of disruptive behaviours after three months~\cite{shaw2020effects}. 

These tools bolster support systems, self-efficacy, and social connection; gaps persist around relationship ambivalence and early identification of compassion fatigue.

\para{Convergent Findings.} Both literature and interviews highlight that online informational and peer platforms strengthen support systems, self-efficacy, and emotional wellness. Studies show that asynchronous forums such as Reddit's r/Alzheimers and moderated programs like Tele-STAR and Talking Point normalize distress and foster social connection~\cite{pickett2024social, lindauer2018took, mckechnie2014effectiveness,kaliappan2025online}. Caregivers similarly described turning to online and peer spaces to \textit{``learn from other people''} and feel \textit{``closer''} to a community (P1). 
These interactions helped manage loneliness, relationship strain, and disruptive behaviors, echoing findings that structured education (e.g., MOOC, FamTechCare) enhances confidence and caregiving competence~\cite{eccleston2019building, shaw2020effects,bhat2023we}. A recurrent toolkit included YouTube channels (e.g., Natalie Edmonds), the Alzheimer's Association website, Reddit's dementia subforums, podcasts, and, for some, AI chatbots. Many anchored their searches in institutionally trusted sources to ensure reliability---Mayo Clinic and the Alzheimer's Association were primary references (P13, P5)---while Wikipedia sometimes supplemented quick fact-checks (P23). Others blended crowd wisdom with structured media such as therapy podcasts or short-form videos for brief, digestible learning (P22, P15). Reddit was valued for its anonymity and normalization function, helping caregivers gauge \textit{``what's normal''} in disease progression (P10) and crowdsource tactics for attention-seeking aggression that \textit{``felt overwhelming''} until support arrived from \textit{``others facing similar challenges''} (P23). Engagement styles ranged from active participation to passive observation; one caregiver described being \textit{``more of a 'troller' than an active contributor''} (P13). Older platforms also persisted: a caregiver used Yahoo Groups for FTD to connect with peers and access information about clinical trials, though access required a physical computer (P2). These experiences collectively reinforce that online communities and multimedia education extend the reach of formal support, fostering connection, practical skill-building, and shared understanding across geographically dispersed caregivers. 

\para{Complementary.} Literature emphasizes that structured, guided interventions reduce reactivity and caregiver strain~\cite{shaw2020effects}, while interviews revealed grassroots adaptations---short YouTube clips and podcasts used for \textit{``digestible learning''} (P13, P15, P22) and AI chatbots providing \textit{``night-time advice''} and synthesized next-step suggestions (P15, P20). Others preferred multimodal consumption: periodic podcasts to \textit{``learn more about dementia''} (P1, P15), brief videos over text-heavy archives (P13), or partner-led medication research online (P24). These behaviors exemplify adaptive information seeking---an evolving blend of expert resources, peer validation, and just-in-time AI mediation. Together, they portray caregivers as active navigators---blending expert, peer, and AI-mediated strategies to preserve emotional balance and mental health within an evolving digital ecosystems. While literature cautions against misinformation and uneven content quality on unmoderated forums~\cite{coulson2024examining, treadgold2025quality}, participants revealed a more complex ambivalence. P24 avoided posting about their own mental health, calling it \textit{``too broad''} among \textit{``100,000 strangers''}, while P2 struggled to find trustworthy or practical financial guidance. Rather than outright distrust, these accounts reveal a tension between accessibility and credibility: caregivers valued openness and immediacy but questioned emotional safety and authority within large, unmoderated spaces. This underscores an unresolved gap between the informational abundance of digital ecosystems and caregivers' need for tailored, trustworthy, and psychologically secure engagement.


\begin{takeaway}
    \textbf{\faLightbulbO\ Takeaway:} Online informational and peer platforms expand caregiver's access to knowledge, emotional validation, and practical problem-solving. This strengthens self-efficacy and social connection. Yet, wide variation in content quality and the absence of professional moderation may introduce risks around misinformation and emotional safety. Therefore, caregivers blend trusted expert sources with peer wisdom and short-form, just-in-time learning---highlighting both the value and the fragility of digital ecosystems in meeting evolving informational needs.
\end{takeaway}

\subsubsection{Social and Emotional Support Technologies (e.g., social media, online support forums)}
\label{sec:social-emotional-tech}
Asynchronous peer forums and social platforms provide safe spaces for emotion sharing and timely advice, mitigating isolation~\cite{kaliappan2025online,pickett2024carevirtue,saha2020causal,kim2023supporters}.
\begin{itemize}
\item \textbf{Emotional expression:} On \textit{Forum.hr}, 32.5\% of posts voiced guilt, loneliness, or helplessness~\cite{erdelez2019online}.
\item \textbf{Responsiveness:} On Dementia Support Forum, 96\% of direct requests were answered the same day (97\% overall)~\cite{coulson2024examining}.
\end{itemize}

A scoping review (19 studies) found \textit{``all studies''} reported benefits of online support groups and social support formation~\cite{daynes2023online}. A 2024 survey ($N=172$) echoed the value but flagged time and e-literacy barriers~\cite{yin2024perceptions}. Embodied modalities: telepresence robots reduced caregiver burden ($p=0.008$), also supporting continuity of care, easing stress and guilt, and complementing in-person visits.~\cite{hung2025impact}; companion robots (e.g., PARO) reduced agitation (meta-analysis, 7 RCTs; $N=214$), indirectly easing carers' stress around disruptive behaviours and sustaining emotional wellness~\cite{lu2021effectiveness}. Short-form platforms (e.g., TikTok) enable narrative sharing and humor as coping, fostering solidarity and reducing stigma~\cite{johnson2022s}. Friendsourcing taps trusted personal networks but suffers variable expertise and potential fatigue~\cite{wilkerson2018friendsourcing}. Overall, these tools bolster support systems, reduce isolation, and improve emotional wellness (with secondary gains for self-efficacy and burnout); gaps remain in proactive relationship maintenance and safeguards against compassion fatigue and digital exclusion.

\para{Convergent Findings.} Both literature and interviews show that peer forums and social platforms mitigate social isolation and loneliness, bolster support systems, and sustain emotional wellness. High rates of emotional expression and same-day responses in dementia forums~\cite{erdelez2019online, coulson2024examining} align with the caregiver's accounts of Reddit as \textit{``a group of people who have the exact same problem,''} offering \textit{``thoughtful''} replies and the reassurance that \textit{``You aren't alone''} (P24, P11; also P19, P22, P21). Caregivers described posting when overwhelmed and receiving supportive comments and resources (P20), learning concrete coping strategies from other's experiences that eased worry (P12), and even \textit{``using Reddit to help others''} as a means of stabilizing their own mindset (P20). Practical peer tips (e.g., bed-fall prevention) translated into action (P12), reinforcing self-efficacy---consistent with reviews that online groups reliably build social support~\cite{daynes2023online}. 
Beyond Reddit, platforms such as CaringBridge and CaringZone enabled update sharing and coordination of practical help that doubled as emotional scaffolding (P12). Broader social networks---LinkedIn for like-minded ties (P6) and Facebook for family updates (P13)---further expanded circles of connection, while several caregivers expressed a desire for \textit{``more'''} dedicated community features (P18). Communication technologies also enhanced presence and belonging: video calls and Alexa-mediated check-ins reduced loneliness and helped others understand the caregiving context (P12, P10), while simple tools such as email and texting maintained everyday proximity to friends and neighbors (P11). 

\para{Complementary Findings.} Interviews also expanded existing frameworks by illustrating how social and emotional technologies extend beyond peer forums to embodied and mindfulness-based modalities. While prior work documents that telepresence tools and companion robots reduce agitation and caregiver burden~\cite{hung2025impact, lu2021effectiveness}, participants described adjacent practices such as routine video calls and Alexa check-ins (P12, P10) that served similar regulatory and connective purposes. Short-form and narrative platforms promoted solidarity and meaning-making~\cite{johnson2022s}, resonating with accounts of \textit{``helping others''} on Reddit as a mood-stabilizing act (P20). Mindfulness and light-touch wellness applications---Headspace and Calm (P1, P23, P11), or Peloton's yoga-meditation routines (P18)---supported emotional balance and daily structure. Together, these findings illustrate that caregivers integrate social, communicative, and mindfulness technologies as flexible coping systems that reduce isolation and reinforce resilience, while remaining aware of the need for moderation and safeguards against digital overuse~\cite{daynes2023online, yin2024perceptions}. Population-level studies often flag time and e-literacy barriers~\cite{yin2024perceptions}, while participants in this study emphasized a different challenge: risks emerging after connection. Some stepped back from social media to avoid addictive engagement or distressing content and voiced concern about TikTok's mental-health effects (P4). Likewise, literature notes fatigue and uneven expertise in friendsourcing~\cite{wilkerson2018friendsourcing}, yet caregivers mostly reported positive, reliable exchanges within family-and-friend networks (P13). These accounts shift the focus from \textit{``access''} to safeguarding compassion fatigue and relationship management over time.



\begin{takeaway}
    \textbf{\faLightbulbO\ Takeaway:} Social and emotional support technologies play an important role in reducing isolation, facilitating emotional expression, and offering timely guidance. However, their benefits are tempered by barriers related to time, digital literacy, and uneven safeguards against compassion fatigue or overreliance. 
\end{takeaway}

\subsubsection{Task Management Tools (e.g., calendars, medication apps)}
\label{sec:task-management-tools}
Trials show that task-management tools buffer key stressors for dementia caregivers. Partner in Balance, an eight-week blended self-management program, increased caregivers' self-efficacy~\cite{bruinsma2021tailoring}. A 12-month Care Ecosystem RCT reduced depression and burden via a remote dashboard for task structuring, medication review and tailored reminders~\cite{possin2019effect}. CareHeroes combined medication-tracking alerts with brief mood/burden checks, improving perceived organizational control and confidence~\cite{brown2016careheroes}. Olera.care couples planning with a curated service directory; usability was near-maximal and perceived competence increased~\cite{fan2024olera}. Self-efficacy and basic self-care are consistently supported. Relationship gains appear mainly when tools link to clinicians or vetted services; compassion fatigue and chronic loneliness remain gaps. Such apps help caregivers strengthen skills, reduce stress, manage negative emotions, build ties and practice self-care~\cite{shen2025perspectives}. Pair scheduling engines with affect-aware analytics and peer matching, and trigger brief evidence-based exercises at reminder time to turn cues into emotionally intelligent micro-interventions.

\para{Convergent Findings.} Trials show that scheduling and medication-support tools boost self-efficacy, reduce depression and burden, and scaffold self-care through routines and reminders~\cite{bruinsma2021tailoring, possin2019effect, brown2016careheroes, fan2024olera, shen2025perspectives}. The interviews strongly echoed these themes. Caregivers described building practical \textit{``task stacks''} to structure daily care, combining calendars, timed prompts, and lightweight aids to keep essentials on track. Several anchored their day with schedulers---Google Calendar and alarms to manage tasks and appointments while staying organized via simple routines (P13), and a virtual assistant to handle scheduling alongside progress logs kept in Excel/Power~BI (P6); others noted calendar-integrated coordination inside CaringBridge/CaringZone to keep caregiving schedules coherent (P12). Medication adherence was scaffolded by toolchains that paired device and reminder layers: a Wi-Fi pill dispenser (Hero) to manage doses (P10), smartwatch time cues to ensure a parent took meds on schedule (P07), and app-based alerts \textit{``to remind myself of the medication''} (P6), with one caregiver using a single app during heavy workloads to queue medication, caregiving tasks, and self-care---and to auto-notify family to check in, which helped prevent overwhelm (P12). For self-maintenance within the same loop, participants leaned on reminder-first wellness apps---Headspace functioning \textit{``like a physical diary''} for prompts (P1) and mobile reminders to take breaks, exercise, and refocus on self-care (P6)---so that care and self-care cues lived in one timetable (P12). Some assembled minimal, glanceable kits to reduce friction at the point of action: a dashboard display to keep key tasks visible and a simplified phone to streamline essential task communications, alongside basic timekeepers and a pill dispenser, albeit with mixed success across components (P05). Overall, participants emphasized simple, timely cues and calendar-native workflows---rather than complex systems---as the most reliable way to translate intention into completion (P13, P6, P10, P12, P07, P05, P1).

\para{Complementary Findings.} Interviews point to lightweight ways task tools can reinforce the support system and emotional wellness: reminder-time family notifications (P12) operationalize coordinated micro-support, aligning with calls to pair planning with affect-aware analytics and service linkage~\cite{shen2025perspectives}. Extending such stacks (Hero, watch, app; dashboard + simplified phone: P10, P07, P12, P05, P1) with brief, triggered exercises and optional clinician or vetted-service connection (as in Care Ecosystem, CareHeroes, or Olera) could better serve relationship management, fatigue prevention, and guard against loneliness~\cite{possin2019effect, brown2016careheroes, fan2024olera}.


\begin{takeaway}
    \textbf{\faLightbulbO\ Takeaway:} Task-management tools help caregivers stay organized, manage medications, and maintain self-care routines, reducing stress and improving confidence. Their potential is greatest when reminders are simple and timely, and when planning tools link to clinical or social support to address longer-term emotional needs.
\end{takeaway}

\subsubsection{Hybrid AI-Driven Platforms and Chatbots}
\label{sec:hybrid-ai}
Conversational agents act as multimodal co-caregivers that screen, educate, and empathize in natural language.
A 2024 review of ten studies found adaptive support improving caregiver effectiveness and well-being~\cite{borna2024artificial}.
In pilot deployment, CareHeroes' chatbot was associated with lower depression at 3 months ($t$=2.03, $p$=0.03)~\cite{ruggiano2024evidence}.
Detection-oriented designs are emerging: HIGEA weaves Zarit-Burden items into Telegram chats and reported early usefulness and effectiveness~\cite{castilla2022higea}.
Chatbots support mood check-ins, psycho-education, and companionship, yet gaps remain: limited tailored coping exercises, sparse clinician escalation, and little coverage of relational micro-skills or anticipatory-grief guidance.
Scaling beyond pilots and showing long-term burn-out reduction are open research needs.


\para{Convergent Findings.} Literature positions chatbots as multimodal co-caregivers that screen, educate, and empathize, improving caregiver effectiveness and wellbeing~\cite{borna2024artificial,weng2026blessing}. The caregivers, during the interviews, similarly described turning to hybrid, AI-enabled tools for decision support, emotional regulation, and accessible companionship. P15 reported using \textit{``AI for [the] care recipient's condition for emergency to resolve the condition''} and querying an AI chatbot at night for timely guidance, while P20 emphasized ChatGPT's directive utility---\textit{``converge all my answers in one direction,''} offer advice, and surface next steps---alongside mood tracking (\textit{``track the depression of a day''}) and emotion logging to monitor strain. Beyond text chat, P14 found AI-based tools and virtual-reality therapy effective for stress relief and burnout prevention, and P1 used a chatbot (Headspace) for mindfulness prompts. These experiences align with pilot evidence showing reduced depression with caregiver-facing chatbots~\cite{ruggiano2024evidence} and mirror detection-style burden prompts in systems such as HIGEA~\cite{castilla2022higea}. Across accounts, participants described gains in self-efficacy, support system strength, relationship management, and emotional wellness (P1, P14, P15, P20).


\para{Complementary Findings.} Calls to scale beyond pilots and demonstrate long-term burnout reduction~\cite{borna2024artificial, ruggiano2024evidence} are complemented by caregiver's hybrid practices. Participants described feasible workflows that combine AI chat for triage (P15), directive guidance and emotional logging (P20), and adjunct calming routines such as VR or structured mindfulness (P14, P1). These hybrid stacks strengthened support without adding burden, showing how AI can slot into existing routines. At the same time, the privacy, empathy, and escalation map~\cite{shi2025mapping} highlights design levers---clear escalation pathways, caregiver-tailored prompts, and emotionally attuned responses---that participants said would meaningfully enhance self-care and emotional wellness over time (P1, P2, P4, P14, P15, P20). Together, the data portray caregivers as active appropriators of AI-enabled tools while pointing to the need for systems specifically designed around the caregivers' emotional, relational, and situational realities. A GPT-4o probe surfaced frictions around privacy, empathy depth, and escalation pathways~\cite{shi2025mapping}. Participants similarly flagged fit gaps: P4 noted awareness of well-being apps but pointed to the absence of caregiver-specific emotional-support features, and P2 viewed AI's primary value as validation, signaling limits in tailored coping and weak safeguards for compassion-fatigue escalation. Accessibility also varied widely: P10 found the Echo Show indispensable for virtual visits that supported lip-reading, whereas P18 reported that Alexa was unusable for their person with dementia---turning instead to a mobile companion dog for acceptable comfort. These differences underscore uneven utility amid communication barriers and enduring disruptive behaviours, and they reveal gaps in current AI system's adaptability to real-world caregiving contexts (P2, P4, P10, P18).


\begin{takeaway}
    \textbf{\faLightbulbO\ Takeaway:} AI-driven platforms and chatbots are emerging as multimodal co-caregivers that provide screening, guidance, emotional validation, and companionship. Early work exhibit reductions in depression and improved caregiver effectiveness, yet gaps persist in tailored coping support, clinician escalation, and relational micro-skill guidance. Their promise lies in integrating empathetic dialogue with adaptive triage and longer-term burnout prevention, which are capabilities that require design maturity and evaluation beyond pilot deployments.
\end{takeaway}

\begin{table}[t]
\centering
\footnotesize
\sffamily
\caption{Comparative summary of technologies and corresponding mental health coverage.}
\begin{tabular}{p{0.15\columnwidth}p{0.24\columnwidth}p{0.27\columnwidth}p{0.24\columnwidth}}
\textbf{Technology} & \textbf{Distinctive features} & \textbf{Capabilities} & \textbf{Persistent blind-spots}\\
\toprule
\rowcollight
Monitoring tools (sensors, trackers) &
Objective, continuous data; real-time alerts; occasional webcam ACT & 
Self-care; burn-out / compassion-fatigue prevention; disruptive-behavior management; emotional wellness &
Little for relationship management, social isolation or anticipatory grief; raw data \(\rightarrow\) limited coaching \\

Informational resources (MOOCs, forums) &
On-demand knowledge from unmoderated forums to structured tele-education &
Support systems; self-efficacy; social isolation buffering; emotional wellness; some compassion-fatigue relief &
Variable accuracy; scarce content on relationship management or early burn-out signs \\ 
\rowcollight
Social \& emotional support tech (forums, robots, video-ACT) &
Empathy-centred interaction; some embodied presence via robots &
Support systems; loneliness reduction; emotional wellness; telepresence cuts burn-out; companion robots calm disruptive behaviours &
Light on compassion-fatigue interventions and fine-grained relationship coaching; access inequities \\

Task-management tools (calendars, dashboards) &
Schedule, track, share tasks; some embed mood check-ins / clinician chat &
Self-efficacy; self-care; occasional burn-out and relationship gains; indirect emotional wellness &
Neglect loneliness, compassion fatigue and disruptive behaviours; risk of notification fatigue \\ 
\rowcollight
Hybrid AI platforms (LLM chatbots, avatars) &
Conversational agents that screen, educate and companion; some human escalation &
Support-system expansion; self-efficacy; self-care; burn-out detection; emotional wellness; loneliness reduction with avatars &
Shallow empathy; few tactics for relationship conflict, anticipatory grief or sustained compassion-fatigue recovery \\
\bottomrule
\end{tabular}
\label{tab:tech-comparative}
\end{table}
\section{Discussion}\label{sec:discussion}



\edit{We organize the Discussion around the two RQs that guided this study. First, we discuss how the taxonomy clarifies the mental health needs of informal AD/ADRD caregivers. Second, we examine how existing technology classes align with these needs and where important gaps remain. We then outline broader design and implementation implications for caregiver-centered technologies and discuss the limitations of this work.}

\subsection{\edit{RQ1: Mental Health Needs of Informal AD/ADRD Caregivers}}
\edit{Towards RQ1, our findings show that informal AD/ADRD caregiver mental health needs extend beyond the commonly used construct of caregiver burden. The taxonomy identifies eight interrelated domains: enduring disruptive behaviors, navigating support systems, building self-efficacy, managing relationships, preventing and mitigating compassion fatigue, practicing self-care, preventing burnout, and regaining emotional balance. These domains show that caregiver mental health is shaped not only by care-recipient symptoms but also by fragmented support systems, family dynamics, moral responsibility, loss of reciprocity, and cumulative emotional exhaustion.}

\edit{A key insight from this synthesis is that caregiver distress unfolds across interacting layers. External stressors, such as disruptive behaviors, care resistance, wandering, sleep disruption, and poorly coordinated services, increase caregiving demands. Relational resources, including family support, communication, shared responsibility, and self-efficacy, can buffer these demands, but may also become sources of strain when responsibilities are unevenly distributed or when family members deny the severity of the condition. Over time, these pressures can contribute to internal depletion, including compassion fatigue, burnout, guilt, anxiety, grief, and emotional imbalance.}

\edit{This layered structure helps explain why caregiver mental health cannot be fully captured by a single burden score. Two caregivers may report similar burden levels but have different underlying needs: one may be struggling with disruptive behaviors and sleep loss, while another may be experiencing family conflict, guilt, or loss of confidence. Distinguishing these domains is therefore important for clinical assessment and for designing technologies that respond to specific caregiver needs.}

\subsection{\edit{RQ2: Alignment and Gaps in Current Caregiver Technologies}}

\edit{Towards RQ2, our findings show that current technologies address some caregiver needs more effectively than others. Monitoring tools and task-management systems are relatively well aligned with safety, organization, medication adherence, and routine support. Informational resources and online peer communities support self-efficacy, social connection, and emotional validation. Hybrid AI platforms and chatbots introduce new possibilities for just-in-time guidance, psychoeducation, mood check-ins, and companionship. However, these technologies remain less aligned with deeper relational and emotional needs, including relationship management, compassion fatigue, anticipatory grief, and long-term burnout prevention.}

\edit{Across technologies, the central misalignment is that many tools are designed around discrete caregiving tasks rather than the evolving emotional and relational experience of caregiving. Monitoring systems can detect wandering or nighttime disruption, but rarely translate these signals into caregiver capacity-building, emotional support, or respite planning. Task-management tools help organize care responsibilities, but often overlook the family dynamics and emotional meanings attached to those tasks. Informational resources provide scalable knowledge, but may lack personalization, moderation, or support for complex emotional situations. AI-based tools show promise, but require stronger safeguards around privacy, empathy depth, escalation, and caregiver-specific adaptation.}

\edit{Together, these findings suggest that caregiver technologies should move beyond information delivery, reminders, and surveillance alone. Caregivers need systems that are stage-sensitive, emotionally responsive, relationally aware, and able to connect caregivers to human support when automated assistance is insufficient.}

\subsection{Design Implications}

 Despite an explosion of digital tools for dementia caregiving, fundamental misalignments persist because the field remains fragmented across nursing, behavioral medicine, human–computer interaction, and AI. Each domain emphasizes different outcomes (e.g., burden reduction, usability, engagement, or accuracy) leading to conceptual drift. Product development cycles also reward short-term engagement and safety alerts rather than long-term emotional adaptation. Finally, evaluation studies are often brief, missing the evolving nature of caregiving. A shared taxonomy and alignment matrix can help bridge these silos by clarifying constructs, outcomes, and stages of need.

\subsubsection{Moving Beyond Surveillance: Integrate Monitoring With Caregiver Capacity-Building}

Caregiving needs are highly stage-dependent. Early in the journey, caregivers seek orientation, medical understanding, and planning support. Mid-stage caregiving demands routine optimization, respite scheduling, and boundary setting. Late-stage caregiving shifts toward grief preparation, safety monitoring, and emotional endurance. Technologies that adapt dynamically to these transitions can deliver tailored content and micro-interventions at the right moment, rather than static reminders that quickly lose relevance.

Monitoring technologies---from actigraphy to ambient sensors and GPS trackers---reduce anxiety, improve sleep, and support safety. Yet our results show a key gap: these systems mainly detect and alert, but rarely build caregiver skills or relational support.
Future designs should link sensing to coaching. When nocturnal agitation is frequent, systems could (with consent) trigger micro-interventions such as relaxation guides, peer nudges, or respite prompts. Dashboards should not only log data but contextualize trends---e.g., linking sleep disruption to burnout or relational strain.

A further need is social integration. Monitoring tools rarely include peer or telepresence features. Embedding optional social layers---for instance, connecting to moderated peer groups when burden scores rise---could reframe monitoring as both safety support and a bridge to resilience.

\subsubsection{Reframe Informational Resources as Scaffolds for Emotional Resilience}

Equity-by-design must be central to future systems. Many caregivers most in need of support, such as those from rural, low-income, ethnoracially minoritized, or low-e-literacy backgrounds, remain systematically underrepresented in both research and technology adoption. Design processes should prioritize multilingual interfaces, plain-language explanations, offline or low-bandwidth functionality, and culturally responsive examples. Evaluations should report subgroup outcomes explicitly rather than collapsing across caregivers with vastly different resources and cultural expectations.

Educational platforms, forums, and tele-mentoring programs provide caregivers scalable knowledge and support. Our results confirm their utility: MOOCs such as \emph{Understanding Dementia} improved competence, forums like \emph{Talking Point} enhanced relationships, and video-feedback interventions reduced reactivity. Yet accuracy is uneven, professional moderation rare, and most content focuses on symptom facts rather than relational or emotional skills.

Caregivers in interviews noted that ``nobody holds your hand'' (P10), with many finding resources only by chance.
 Even strong programs were episodic, addressing isolated concerns but not sustaining needs across the disease course. Informational resources should thus be reframed as scaffolds for emotional resilience, adapting to caregiving stage and stress while embedding reflective exercises, communication tips, and fatigue detection.

Designs must also address trust--reliability tradeoffs. Forums offer empathy yet risk misinformation. Tiered credibility markers---distinguishing peer anecdotes from clinician-reviewed content---and AI-based summarization or fact-checking can mitigate this burden. Continuity features (e.g., progress tracking, revisiting discussions) can transform episodic tools into persistent companions for resilience-building.

\subsubsection{Design Social and Emotional Technologies for Sustained Relational Health}

Relationship health deserves to be treated as a primary outcome of caregiving technologies. Our data showed that sibling inequity, partner strain, and friendship loss were among the strongest predictors of distress.

Peer forums, social platforms, and embodied tools such as telepresence and companion robots reduce loneliness and buffer distress. Literature and our findings confirm: online forums were supportive, robots like PARO reduced agitation, and telepresence eased guilt. Yet two gaps emerged. First, support is reactive and episodic---caregivers seek help only when overwhelmed, while tools seldom detect early relational strain or compassion fatigue. Second, technologies target individual expression rather than family dynamics, despite sibling inequity, partner strain, and friendship loss being dominant stressors.

Design must shift toward sustained relational health. Future systems could integrate relational sensing and coaching, e.g., detecting resentment in communication logs or prompting family check-ins. AI facilitation may aid conflict-mitigation or fair task division, addressing the ``I'm the only one''(P3) dynamic. Proactive fatigue prevention is also critical: irritability, loss of motivation, or ``snapping'' were frequent early cues. Embedding lightweight check-ins and opt-in referral could prevent escalation.

\subsubsection{Evolve Task-Management Tools into Context-Aware Micro-Interventions}

Task-management platforms---from calendars to medication dashboards---show measurable benefits in reducing burden, boosting self-efficacy, and improving organization. Yet reminders can overwhelm during emotional depletion, and organizational control does not always build relational or psychological resilience~\cite{wrede2021requirements,ohly2023effects}.

Design must therefore move beyond logistics into context-aware micro-interventions. For example, if a medication alert coincides with caregiver fatigue, the app could trigger a brief breathing exercise or peer validation. Scheduling engines could detect burnout indicators and dynamically adjust pacing or nudge for respite. Gains were most evident when tools linked to clinicians or vetted services, suggesting referral engines and family communication boards are critical.

\subsubsection{Reimagine Hybrid AI Platforms as Longitudinal Co-Caregivers}

Conversational agents are promising yet contested. Studies show chatbots can reduce depression, deliver psychoeducation, and offer companionship. Our probe with a GPT-4o system (``Carey'') echoed these benefits but exposed limits: shallow empathy, absent escalation, and weak coverage of relational micro-skills or anticipatory grief. Caregivers valued an always-available interlocutor, but worried about trust, privacy, and dependency.

Hybrid AI must thus evolve into longitudinal co-caregivers. First, they require persistent, consent-driven memory to recall prior exchanges, track emotional states, and personalize across the caregiving journey; without continuity, support remains ``flat and moment-specific.'' Second, AI must adapt to caregiving stages and crises---from education for novices to burnout or grief preparation for long-term caregivers---with tiered escalation to human support.

Equally important is balancing privacy and personalization. Tiered data-sharing can let users choose disclosure levels: lightweight generic coaching versus deeper crisis planning. Finally, AI should link into broader ecosystems of case managers, support groups, or clinics. Positioned as ethically aware, stage-sensitive, and connected, AI platforms can strengthen caregiver resilience without displacing human bonds.

\subsection{Limitations and Future Directions}

In this section, we highlight potential limitations of our study that warrant consideration. We also discuss future related work that could be undertaken to strengthen our cumulative understanding in this field. Firstly, our qualitative sample (N=41 total) was primarily recruited from online social media communities in the U.S., where many AD/ADRD caregivers seek support and share lived experiences~\cite{kaliappan2025online}. 
This could potentially over-represent digitally literate, English-speaking and help-seeking caregivers. 
\edit{All interviews were conducted in English with participants who indicated comfort completing the interview in English. Interviewers used follow-up prompts and clarification questions to support participants’ expression and ensure that meanings were understood. No translation was involved. We acknowledge that this may have limited the inclusion of caregivers who would have preferred to participate in another language.}
Future studies could prioritize the recruitment of rural, low-income, ethnoracially minoritized, and non-English-speaking populations to examine how structural inequities shape mental health needs and technology access. Cross-cultural comparative research is especially needed given divergent caregiving norms and support infrastructures across global contexts. 

Second, while our sample included spousal and adult-child caregivers across disease stages, subgroup analyses were limited by sample size. Prior work demonstrates that spousal caregivers---particularly wives---report lower self-efficacy and greater strain~\cite{depp2005caregiver}, while caregiving-demands shift substantially from early (orientation, planning) to middel (behavioral crises) to late (grief, safety) stages. Longitudinal studies tracking caregivers across disease progression are essential to refine stage-sensitive intervention designs and capture dynamic need-evolution rather than cross-sectional snapshots.

Third, while triangulation across literature and interviews identified points of convergence, contradiction, and complementarity, the analysis does not imply causal relationships. Instead, it should be interpreted as a conceptual mapping of psychosocial and relational dynamics. Moreover, the technology-mapping exercise primarily drew from visible research and commercial prototypes, which may bias interpretations toward formalized systems rather than grassroots or culturally situated support mechanisms. Future work should explore these contextual nuances and examine how relational and technological dynamics co-evolve across diverse caregiving settings. 

\edit{Finally, our literature review is that our search strategy used ``caregiver'' as the primary search term and did not include ``care partner'' as a separate keyword. We selected ``caregiver'' because it remains the dominant indexing term across the interdisciplinary literature reviewed in this study. However, we recognize that ``care partner'' is increasingly used in dementia research as a more person-centered term. As a result, our review may have missed studies that only used ``care partner'' rather than ``caregiver.'' Future reviews should explicitly include ``care partner,'' ``family carer,'' and related terms to better reflect evolving terminology in dementia care research.} 
\section{Conclusion}

Caregivers of individuals with Alzheimer's disease and related dementias provide the foundation of long-term care, yet they experience substantial mental health burdens that remain inadequately characterized and supported. This study addressed critical gaps by developing a systematic taxonomy linking caregiver mental health needs to technology-based interventions. Through triangulation of an interdisciplinary literature review and qualitative studies with 41 caregivers, we disaggregated the construct of ``caregiver burden'' into eight empirically grounded domains: managing disruptive behaviors, navigating support systems, building self-efficacy, sustaining relationships, preventing compassion fatigue, practicing self-care, avoiding burnout, and maintaining emotional wellness. Our findings revealed systematic misalignments between caregiver needs and current technologies. Insights from this study yield actionable design directions: integrate monitoring with capacity-building, embed emotional scaffolding within informational content, target family dynamics through transparent task allocation and conflict detection, deliver context-aware micro-interventions adapted to caregiver state, and implement AI with persistent memory and transparent escalation. By systematically characterizing caregiver's multidimensional needs and technology gaps, this study establishes a roadmap for adaptive, personalized, and holistic interventions that support caregivers throughout the entire caregiving journey, benefiting both caregivers and care recipients.


\backmatter





\bmhead{Acknowledgments}

This work was supported in part by the National Institute on Aging of the National Institutes of Health under Award Number P30AG073105 and the Jump ARCHES endowment through the Health Care Engineering Systems Center at the University of Illinois, and the OSF Foundation.










\begin{appendices}







\end{appendices}


\bibliography{0paper}


\end{document}